\documentclass[aps,pra,floatfix,superscriptaddress,showpacs,twocolumn,nofootinbib,longbibliography]{revtex4-1}
\usepackage{amsthm}
\usepackage{amsfonts}
\usepackage{amsmath}
\usepackage{amssymb}
\usepackage{mathtools}
\usepackage{graphicx}
\usepackage{color}
\usepackage{accents}
\usepackage{bbm}
\usepackage{bm}
\usepackage{multibib}
\usepackage[hidelinks]{hyperref}
\hypersetup{
     colorlinks   = true,
     citecolor    = blue,
     linkcolor    = blue
}
\usepackage{matlab-prettifier}
\usepackage{overpic}

\newcommand{\ignore}[1]{}
\newcommand{\nobibentry}[1]{{\let\nocite\ignore\bibentry{#1}}}

\newcommand{\bibfnamefont}[1]{#1}
\newcommand{\bibnamefont}[1]{#1}
\newcommand{\average}[1]{\langle#1\rangle}
\newcommand{\ket}[1]{\left\vert#1\right\rangle}
\newcommand{\bra}[1]{\left\langle#1\right\vert}

\definecolor{darkgreen}{RGB}{50,190,50}

\begin{document}
\title{On super additivity of Fisher information in fully Gaussian metrology}
\author{Javier Navarro}
\email{jnavarro@bcamath.org}
\affiliation{Basque Center for Applied Mathematics (BCAM), Alameda de Mazarredo 14, 48009 Bilbao, Spain}
\affiliation{Department of Physical Chemistry, University of the Basque Country UPV/EHU, Apartado 644, 48080 Bilbao, Spain}
\author{Simon Morelli}
\affiliation{Vienna Center for Quantum Science and Technology (VCQ), Atominstitut, TU Wien, Stadionallee 2, 1020 Vienna, Austria}
\author{Mikel Sanz}
\affiliation{Basque Center for Applied Mathematics (BCAM), Alameda de Mazarredo 14, 48009 Bilbao, Spain}
\affiliation{Department of Physical Chemistry, University of the Basque Country UPV/EHU, Apartado 644, 48080 Bilbao, Spain}
\affiliation{IKERBASQUE, Basque Foundation for Science, Plaza Euskadi 5, 48009 Bilbao, Spain}
\affiliation{EHU Quantum Center, University of the Basque Country UPV/EHU, Bilbao, Spain}
\author{Mohammad Mehboudi}
\email{mohammad.mehboudi@gmail.com}
\affiliation{Vienna Center for Quantum Science and Technology (VCQ), Atominstitut, TU Wien, Stadionallee 2, 1020 Vienna, Austria}

\begin{abstract}
    Famously, the quantum Fisher information---the maximum Fisher information over \textit{all} physical measurements---is additive for independent copies of a system and the optimal measurement acts locally. 
    We are left to wonder: does the same hold when the set of accessible measurements is constrained? Such constraints are necessary to account for realistic experimental restrictions. Here, we consider a \textit{fully Gaussian scenario} focusing on \textit{only} Gaussian measurements. We prove that the optimal Gaussian measurement protocol remains local, if the information is encoded in either the displacement or the covariance matrix.
    However, when the information is imprinted on both, this no longer holds true: we construct a simple global Gaussian measurement where the Fisher information becomes super additive. These results can improve parameter estimation tasks via feasible tools. Namely, in quantum optical platforms our proposed global operation requires only passive global operations and single mode Gaussian measurements.
    We demonstrate this in two examples where we estimate squeezing and losses. While in the former case there is a significant gap between the Fisher information of the optimal Gaussian measurement and the quantum Fisher information for a single copy, this gap can be reduced with joint Gaussian measurements and closed in the asymptotic limit of many copies.
\end{abstract}
\maketitle 

{\it Introduction---}
The quantum Fisher information (QFI) is a central object in the statistics and dynamics of quantum systems~\cite{Scandi_2025,GU2010,Zanardi07,CamposZanardi07}, where it defines a metric on the quantum state space~\cite{braunstein1994statistical,Paraoanu98,petz2001extendingfishermetricdensity,slater2003brodyhughstonfisherinformationmetric,Safranek_2017}, and can be used to find upper bounds on the speed of quantum processes~\cite{PhysRevX.12.011038,PhysRevA.85.052127,PhysRevX.6.021031,PhysRevLett.110.050402,PhysRevA.97.022109}. Probably it is best known to the reader for its role in parameter estimation, where it shows up in both frequentist~\cite{rao1992information,cramer1999mathematical,frieden2004science,braunstein1994statistical,paris2009quantum,Toth_2014,Braunstein1996,DEMKOWICZDOBRZANSKI2015345} and Bayesian [quantum] Cram\'er-Rao bounds ([Q]CRB) \cite{van2004detection,Gill1995,Li2018}, prior design~\cite{Jeffry1946}, and asymptotic posterior distribution~\cite{Vaart1998,Hartigan1983,Cam1986}. The QFI has fundamental importance in determining the ultimate precision scaling in quantum metrology in the asymptotic limit of large data~\cite{PhysRevLett.96.010401,GiovannettiLloydMaccone2011,PhysRevLett.98.090401}. Nonetheless, when it comes to practicality, its relevance in quantum metrology  can be questioned from two aspects: (i) not always one has access to large data and (ii) not always one has access to the optimal measurements. A more formal definition is in order: 
Take a quantum system with $\rho(\theta)$ denoting its density operator acting on a Hilbert space ${\cal H}$. Let $M$ be a measurement, or a positive operator valued measure (POVM), with elements $\{M_x\}_x$ also acting on ${\cal H}$. In this setup the Born rule determines the observation probabilities $p(x|\theta) = {\rm Tr}[M_x \rho(\theta)]$.
The Fisher information (FI) is defined as the expected value of the observed information~\cite{Fisher22}
\begin{equation}
    {\cal F}(\rho(\theta);M) \coloneqq \int dx\ p(x|\theta) \left[ \partial_{\theta} \log p(x|\theta)  \right]^2,
\end{equation}
where we take the outcome space to be continuous, but in principle it can also be discrete.
In turn, the QFI is the maximum FI over all possible measurements ${\cal F}^Q(\rho(\theta)) \coloneqq \max_{M} {\cal F}(\rho(\theta);M)$.

The relevance of the FI and the QFI in the frequentist approach to parameter estimation is easier to express and thus we will take this approach here. For any unbiased estimator ${\hat \theta}$ (i.e., $\average{{\hat \theta}} = \theta$) the mean square error ${\rm MSE} \coloneqq \average{(\hat \theta(x) - \theta)^2}_{p(x|\theta) }$ is lower bounded~\cite{rao1992information,cramer1999mathematical}
\begin{align}
    {\rm MSE} \geq \frac{1}{\nu {\cal F}(\rho(\theta); M)} \geq \frac{1}{\nu {\cal F}^{\rm Q}(\rho(\theta))},
\end{align}
where $\nu$  
is the number of measurement repetitions (iid).
Note that (i) the first inequality, known as the CRB, holds only for unbiased estimators and theoretically its saturation can be guaranteed in the asymptotic limit of $\nu \gg 1$. (ii) The second inequality, known as the QCRB, can be saturated, in theory, by definition of the QFI \cite{braunstein1994statistical,paris2009quantum,Toth_2014,DEMKOWICZDOBRZANSKI2015345}. However, the optimal measurement that is required for its saturation can be experimentally out of reach. 
As such, taking the QFI as the figure of merit is justified only when these two points are properly addressed.

Exactly these two points motivate the study of fully Gaussian metrology: Encoding of physical parameters on Gaussian systems by means of Gaussian quantum channels is a common and approachable practice in quantum optical platforms \cite{Caruso2008,KlauderSkagerstam1985,AndersenGehringMarquardtLeuchs2016} and Gaussian measurements are also fairly available techniques by means of linear optics and homodyne detection \cite{Gerry2004,Bachor2019,Genoni_2011,GiovannettiLloydMaccone2011,DEMKOWICZDOBRZANSKI2015345,Oh2019}. Furthermore, the Gaussian statistics that is produced as a result are among rare ones that can saturate CRB with small data \cite{lehmann1998theory,10.5555/151045,kolodynski2014precision}. 

Here, we raise a question within the fully Gaussian framework---see below for a more formal statement: Is the FI of the optimal Gaussian measurement additive in the number of copies (just like the FI of the optimal measurement, i.e., the QFI) or can joint operations surpass this bound? We prove that joint operations can indeed give an advantage if the parameter is encoded on both the first and the second moment. Remarkably, by using such joint Gaussian operations, we can reduce (and sometimes fully close) the gap between the FI of the optimal Gaussian measurement and the QFI. 

{\it Preliminaries---}Take a bosonic system with $k$ modes, with the quadratures $R = [q_1, p_1, \dots, q_k, p_k]^T$ that satisfy the canonical relations $[R_i,R_j] = i\hbar \Omega_{ij}$. Here, $\Omega = \oplus_{j=1}^{k}\left(\begin{smallmatrix}
    0 & 1\\-1 & 0
\end{smallmatrix}\right)$ is the symplectic form and we set $\hbar=1$ in what follows. Suppose that the system's quantum state, represented by the density operator $\rho_{\theta}$ is Gaussian. This implies that it can be fully characterized by its first and second moments denoted by
$d_{\theta} = {\rm Tr}[\rho_{\theta} R ]$, and 
$\sigma_{\theta} = {\rm Tr}[\rho_{\theta} \{(R-d),(R-d)^T\} ]$ respectively. The Heisenberg uncertainty principle dictates that $\sigma_{\theta} + i \Omega \geq 0$~\cite{PhysRevA.49.1567,RevModPhys.84.621}. Likewise, Gaussian measurements on the system, i.e., those that when applied on Gaussian systems produce outcomes with Gaussian distributions, can be fully characterized by a covariance matrix $\sigma^M$ that also respects the uncertainty principle~\cite{Kiukas2013}. The outcome distribution can be written as
\begin{align}\label{eq:Gaussian_meas_pdf}
p(a|d_{\theta},\sigma_{\theta};\sigma^M) = \frac{e^{-\frac{1}{2}(a-d_{\theta}^T)(\sigma_{\theta} + \sigma^M)^{-1}(a-d_{\theta}^T)}}{(2\pi)^k\sqrt{{\rm det} (\sigma_{\theta} + \sigma^M)}}.
\end{align}

The QFI for Gaussian quantum systems is well known and can be related to the moments and their derivatives in a closed formula~\cite{monras2013phase,PinelJianTrepsFabreBraun2013}
\begin{align}
    {\cal F}^{\rm Q}(d_\theta,\sigma_\theta)   & = 2\partial_{\theta} d_\theta^{T} \sigma_\theta^{-1} \partial_{\theta} d_\theta
    \nonumber\\
    & ~+ \frac{1}{2}\langle\hspace{-.6mm}\bra{\partial_{\theta}\sigma_{\theta}} [\sigma_{\theta} \otimes \sigma_{\theta} - \Omega \otimes \Omega]^{-1} \ket{\partial_{\theta}\sigma_{\theta}}\hspace{-.6mm}\rangle,
\end{align}
with $\ket{\bullet}\hspace{-.6mm}\rangle$ being a vectorization of $\bullet$.
On the other hand the FI of a Gaussian measurement $\sigma^M$ is given by \cite{Malag2015,cenni2022thermometry}
\begin{align}\label{eq:FI_Gaussian_main}
{\cal F}^{\rm G}({d_\theta},\sigma_\theta;\sigma^{M})  = {\cal F}^{\rm G}_d({d_\theta},\sigma_\theta;\sigma^{M})  + {\cal F}^{\rm G}_{\sigma}(\sigma_\theta;\sigma^{M}),
\end{align}
where
\begin{align}
  & {\cal F}^{\rm G}_d(d_\theta,\sigma_\theta;\sigma^{M})    = 2\partial_{\theta} d_\theta^{T}[\sigma_\theta + \sigma^{M}]^{-1}\partial_{\theta} d_\theta, \label{eq:FI_displacement} \\
  &{\cal F}^{\rm G}_{\sigma}(\sigma_\theta;\sigma^{M})= \frac{1}{2}{\rm Tr} \left[\left(( \sigma_\theta + \sigma^{M})^{-1} \partial_{\theta} \sigma_\theta \right)^2 \right].\label{eq:FI_2terms}
\end{align}
We can further define the optimal Gaussian measurement and FI as follows
\begin{align}
    \accentset{\ast}{\sigma}^M & \coloneqq \arg\max_{\sigma^M} {\cal F}^{\rm G}(d_\theta,\sigma_\theta;\sigma^{M}), \nonumber \\
    \accentset{\ast}{\cal F}^{\rm G}(d_{\theta}, \sigma_{\theta}) & \coloneqq {\cal F}^{\rm G}(d_\theta,\sigma_\theta;\accentset{\ast}{\sigma}^M). \nonumber
\end{align}
Note that the optimal Gaussian measurement can be always represented by a pure covariance matrix~\cite{cenni2022thermometry}.
We are now ready to formally state the problem.

{\it Statement of the problem---}Take $m$ copies of the state, i.e., $\rho_{\theta}^{\otimes m}$. That amounts to taking a Gaussian system with the displacement $d^{\oplus m}$ and the covariance matrix $\sigma_{\theta}^{\oplus m}$. It is known that the QFI of this state is additive~\cite{Ji2008} (this is true independent of the state's Gaussianity), that is
\begin{align}
    {\cal F}^{\rm Q}(d_{\theta}^{\oplus m} , \sigma_{\theta}^{\oplus m}) = m {\cal F}^{\rm Q}(d_{\theta} , \sigma_{\theta}).
\end{align}
This suggests that global operations among identical copies can not improve the QFI~\cite{GiovannettiLloydMaccone2011}.

When it comes to Gaussian measurements, we know for sure that 
\begin{align}
    \accentset{\ast}{\cal F}^{\rm G}(d_{\theta}^{\oplus m} , \sigma_{\theta}^{\oplus m}) 
    & \geq {\cal F}^{\rm G}(d_{\theta}^{\oplus m} , \sigma_{\theta}^{\oplus m}; (\accentset{\ast}{\sigma}^M)^{\oplus m}) \nonumber\\
    & = m \accentset{\ast}{\cal F}^{\rm G}(d_{\theta} , \sigma_{\theta}),
\end{align}
since $(\accentset{\ast}\sigma^M)^{\oplus m}$ is a feasible Gaussian measurement.
The question is, if/when the inequality is strict. This amounts to finding global Gaussian measurements that improve the FI super additively. In what follows we present our main results, which involve a necessary (but not sufficient) condition for the strict inequality to hold, and then construct examples where global operations can lead to a super additive Gaussian FI.

{\it Result 1: additivity of the Gaussian FI---}When the parameter is encoded either on the displacement or the covariance matrix, the optimal Gaussian FI is additive. In the case of displacement encoding it is well known that the optimal measurement that achieves the QCRB is homodyne detection in a suitable direction, which is a Gaussian measurement. As such, global Gaussian operations cannot further improve the precision. See ~\cite{Personick1971} or Appendix \ref{app:proof} for a proof.

Similarly, we prove that when the parameter is encoded only in the covariance matrix of single mode systems through any Gaussian evolution, global Gaussian operations do not offer an advantage over local ones. The proof is slightly technical, but can be followed by basic knowledge of linear algebra, and thus we present a simple version of it here, and leave a generalization to isothermal models with more than one mode to Appendix~\ref{app:IsothermalCov}. 
\begin{proof}By definition, for any (non-local) pure $\sigma^M$ we have
\begin{align*}
      {\cal F}^{\rm G}(d^{\oplus m},\sigma^{\oplus m}_\theta;\sigma^{M})   =  \frac{1}{2}\operatorname{Tr} \left[\left((\sigma^{\oplus m}_\theta +  \sigma^{M})^{-1} \partial_{\theta}\sigma^{\oplus m}_\theta\right)^2 \right]\\
      %%%%%
      =\frac{1}{2}\operatorname{Tr} \left[\left((\mu I +  \gamma^{M})^{-1} [S^{-1}\partial_{\theta}\sigma_\theta (S^{T})^{-1}]^{\oplus m}\right)^2 \right],
\end{align*}
where in the second line we used the Williamson decomposition \cite{RevModPhys.84.621} $\sigma_\theta  = \mu S S^T  $ for some symplectic transformation $S$ and defined $\gamma^{M}=(S^{-1})^{\oplus m}\sigma^{M}(S^{-T})^{\oplus m}$---we denote $S^{-T} = (S^{T})^{-1}$.
Since $ \gamma^{M} $ is the covariance matrix of a pure measurement it can be diagonalized by a global orthosymplectic $O_G$ \cite{serafini2023quantum}
\begin{align*}
   \gamma^{M} &= O_G(\oplus_{i=1}^mZ_i) O_G^T= O_G\left(\oplus_{i=1}^m \left(\begin{smallmatrix}z_i&0\\0&1/z_i\end{smallmatrix}\right)\right) O_G^T.
\end{align*}
Furthermore, note that $(\mu I + \oplus_iZ_i)^{-1}=\oplus_{i=1}^m{\rm diag}(\tfrac{1}{\mu+z_i},\tfrac{z_i}{\mu z_i+1})\eqqcolon V$. 
%\simo{(I usually avoid fractions in text, maybe $(1/(\mu+z_i),z_i/(\mu z_i+1))$ or at least $(\tfrac{1}{\mu+z_i},\tfrac{z_i}{\mu z_i+1})$.)} 
By further defining $W\coloneqq [S^{-1}\partial_{\theta}{\sigma}(S^{T})^{-1}]^{\oplus n}$ we have
\begin{align}
    {\cal F}^{\rm G}(d^{\oplus m}, \sigma^{\oplus m}; \sigma^{M})  =\frac{1}{2}\operatorname{Tr} \left[\left(V\: O_G^T WO_G\right)^2 \right]\nonumber\\\leq \frac{1}{2}\operatorname{Tr} \left[ V^2 O_G^T W^2 O_G  \right].
    \label{eq:ineq1}
\end{align}
The inequality can be proven straightforwardly. Define ${\tilde W} = O_G^T WO_G$. Note that both matrices $V$ and ${\tilde W}$ are real symmetric matrices. Using ${\rm Tr} [[V,{\tilde W}]^2] = 2 {\rm Tr} [(V{\tilde W})^2] - 2{\rm Tr} [V^2{\tilde W^2}] $, and that $[V,{\tilde W}]^T = -[V,{\tilde W}]$, we have ${\rm Tr}[[V,{\tilde W}]^2] \leq 0$. This proves that ${\rm Tr} [(V{\tilde W})^2] \leq {\rm Tr} [V^2{\tilde W^2}] = {\rm Tr} [(V^2 O_G W^2 O_G)]$. This inequality is tight if $[V, O_G W O_G] = 0$.

To proceed further, we take $\{{W_i}^2\}$ to be the set of eigenvalues of ${W}^2$ sorted descendingly, i.e., $ W_i^2\geq W_j^2$ if $i<j$. Similarly take the eigenvalues of $V^2$, to be $\{V_i^2\}$ descendingly ordered. By using the majorisation theorem, we show in Appendix~\ref{app:proof} that
\begin{align}
      \operatorname{Tr} \left[ V^2  \left(O_G^T W^2O_G \right)  \right] \leq \sum_i V_i^2W_i^2.
      \label{eq:ineq2}
\end{align}
The equality holds when $ O_G$ is a permutation matrix so that it sorts the eigenvalues of $W$ according to $V$. In the particular problem we are studying $W^2= (w^2)^{\oplus m}$ and $V^2=(v^2)^{\oplus m}$ are block diagonal. So an  optimal transformation---that saturates inequalities \eqref{eq:ineq1} and \eqref{eq:ineq2}---is the local transformation $O_G=O_L^{\oplus m}$ that acts in a single block as follows
\begin{align}
     O_L^T w^2O_L = \begin{pmatrix}
         w_1^2&0\\
         0&w_2^2
     \end{pmatrix} \quad \textit{if}\quad v^2 = \begin{pmatrix}
         v_1^2&0\\
         0&v_2^2
     \end{pmatrix}, \\ 
 O_L^T w^2O_L = \begin{pmatrix}
         w_2^2&0\\
         0&w_1^2
     \end{pmatrix} \quad \textit{if}\quad v^2 = \begin{pmatrix}
         v_2^2&0\\
         0&v_1^2
     \end{pmatrix}.  
\end{align}
To sum up, the optimal Gaussian measurement is a local one with the form
\begin{align}\label{eq:opt_local}
    \sigma^M = \oplus_{i=1}^m S O_L\left(\begin{smallmatrix}
        z & 0 \\ 0 & 1/z
    \end{smallmatrix}\right)O_L^T S^T.
\end{align}
\end{proof}

Now that we know the optimal measurement is indeed local, one can ask which specific local Gaussian measurement is optimal?
This amounts to finding the optimal rotation (permutation) $O_L$ and the optimal $z$, 
\begin{align*}
\max_{\{z\} }\sum_{i=1}^2 v_i^2w_i^2
      =\max\{f(w_1,w_2), f(w_2,w_1)\}  
\end{align*}
with $f(w_1,w_2) = \max_{z}w_1^2/(\mu + z)^2 + w_2^2z^2/(z \mu + 1)^2$.

This is a simple optimization problem. For instance, in problems where $w_1=w_2$---example, thermometry, or unitary (symplectic) encoding on a thermal state---if  $\mu\leq1+\sqrt{2}$ the maximum Fisher information is obtained at $z^*\to 0$ which corresponds to homodyne detection. Otherwise, for  $\mu\geq 1+\sqrt{2}$ the optimal value is $z^*=1$ corresponding to heterodyne detection. This result further generalizes those of \cite{cenni2022thermometry} found for Gaussian thermometry.
%%%%%%%%%

{\it Result 2: Super additivity of the Gaussian FI---}Super additivity of the optimal Gaussian FI can be demonstrated by a simple construction. Let us point out an important aspect when the parameter is encoded on both moments, which will then naturally lead to our construction. The FI of a Gaussian measurement \eqref{eq:FI_Gaussian_main} contains two terms. The optimal Gaussian FI requires finding a measurement with covariance matrix $\accentset{\ast}{\sigma}^M$ that maximizes the sum. Generally, the two terms may be optimized separately by using different (incompatible) Gaussian measurements $\accentset{\ast}{\sigma}^M_d \coloneqq \arg\max_{\sigma^M} {\cal F}^{C}_d(d_\theta,\sigma_\theta;\sigma^{M})$ and $\accentset{\ast}{\sigma}^M_{\sigma} \coloneqq \arg\max_{\sigma^M} {\cal F}^{C}_{\sigma}(d_\theta,\sigma_\theta;\sigma^{M})$. Thus, the maximum of the sum is generally less than the sum of the maxima of the individual terms.

\begin{figure}
    \centering
    \includegraphics[width=.9\linewidth]{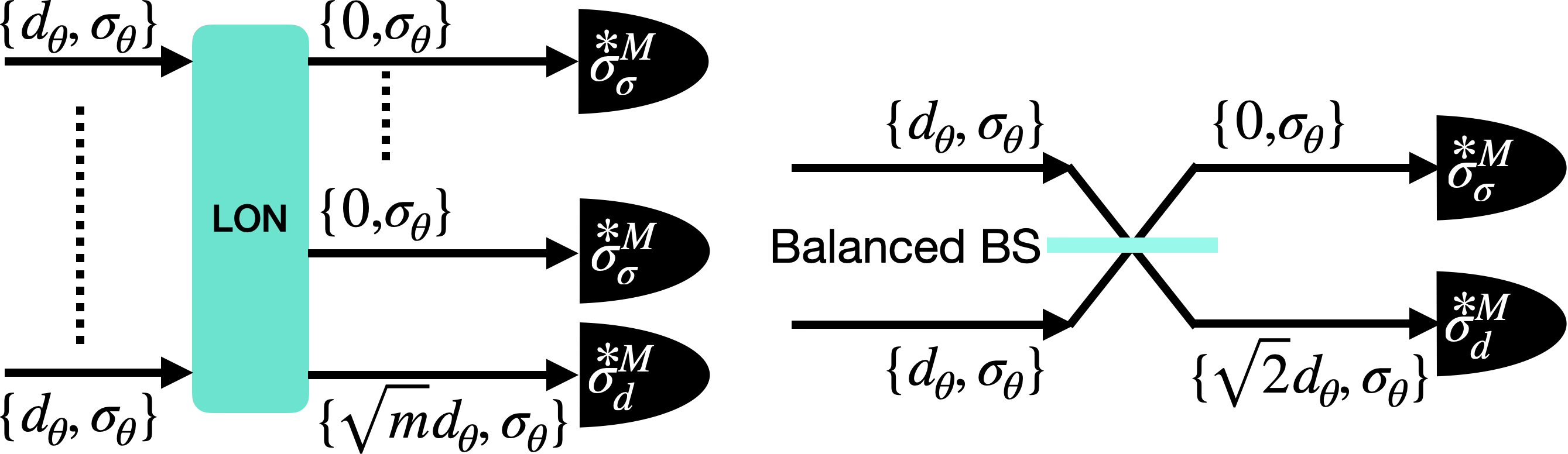}
    \caption{Schematic of our proposed global Gaussian measurement that can demonstrate super additivity of the optimal Gaussian FI. The measurement strategy only require LON and local Homodyne/Heterodyne detection. In case of $m=2$ the LON reduces to a simple balanced beam splitter. }
    \label{fig:LON}
\end{figure}

\begin{align}\label{eq:1mode_bound}
    \accentset{\ast}{\cal F}^{\rm G}(d_{\theta}, \sigma_{\theta}) \leq {\cal F}^{\rm G}_d(d_{\theta}, \sigma_{\theta};\accentset{\ast}{\sigma}^M_d ) + {\cal F}^{\rm G}_{\sigma}(d_{\theta}, \sigma_{\theta};\accentset{\ast}{\sigma}^M_{\sigma} )
\end{align}
Interestingly, this issue can be simply overcome in the many copy regime. As depcited in Fig.~\ref{fig:LON} there always exists a linear optical network (LON) which, crucially, is independent of $\theta$ and makes the following transformation
\begin{align}
    d_{\theta}^{\oplus m} & \overset{\rm LON}{\longmapsto} 0^{\oplus m-1} \oplus (\sqrt{m} d), \nonumber\\
    \sigma_{\theta}^{\oplus m} & \overset{\rm LON}{\longmapsto}\sigma_{\theta}^{\oplus m}.
\end{align}
In other words, the LON shifts  the displacement from all modes to the last mode, while leaving the covariance matrix untouched. The output is therefore \textit{not entangled}. As such, one can perform a Gaussian measurement with covariance matrix $[\accentset{\ast}{\sigma}^M_{\sigma}]^{\oplus m-1} \oplus \accentset{\ast}{\sigma}^M_d$ to obtain
\begin{gather}
    {\cal F}^{\rm G}({\rm LON}\{d_{\theta}^{\oplus m}, \sigma_{\theta}^{\oplus m}\}; [\accentset{\ast}{\sigma}^M_{\sigma}]^{\oplus m-1} \oplus \accentset{\ast}{\sigma}^M_d)\nonumber\\
    =  (m-1){\cal F}^{\rm G}( 0,\sigma_{\theta}; \accentset{\ast}{\sigma}^M_{\sigma}) + {\cal F}^{\rm G}(\sqrt{m}d_{\theta}, \sigma_{\theta}; \accentset{\ast}{\sigma}^M_d)  
    \nonumber\\
    \geq  (m-1){\cal F}^{\rm G}_{\sigma}( \sigma_{\theta}; \accentset{\ast}{\sigma}^M_{\sigma}) + m{\cal F}^{\rm G}_{d}(d_{\theta}, \sigma_{\theta}; \accentset{\ast}{\sigma}^M_d),\label{eq:FI_LON}
\end{gather}
where in the second line we used that the LON does not create correlations among the copies
and in the last line we used that ${\cal F}^{\rm G}(0, \sigma_{\theta}; \accentset{\ast}{\sigma}^M_{\sigma}) = {\cal F}^{\rm G}_{\sigma}(\sigma_{\theta}; \accentset{\ast}{\sigma}^M_{\sigma})$ since the first moment is zero, and ${\cal F}^{\rm G}(\sqrt{m}d_{\theta}, \sigma_{\theta}; \accentset{\ast}{\sigma}^M_d) \geq {\cal F}^{\rm G}_d(\sqrt{m}d_{\theta}, \sigma_{\theta}; \accentset{\ast}{\sigma}^M_d) = m {\cal F}^{\rm G}_d(d_{\theta}, \sigma_{\theta}; \accentset{\ast}{\sigma}^M_d)$---see Eq.~\eqref{eq:FI_displacement}. Note that Eq.~\eqref{eq:FI_LON} suggests that at the limit of $m\gg 1$ we have found a global Gaussian measurement strategy  for which ${\cal F}^{\rm G}({\rm LON}\{d_{\theta}^{\oplus m}, \sigma_{\theta}^{\oplus m}\}; [\accentset{\ast}{\sigma}^M_{\sigma}]^{\oplus m-1} \oplus \accentset{\ast}{\sigma}^M_d) \approx m {\cal F}^{\rm G}_d(d_{\theta}, \sigma_{\theta};\accentset{\ast}{\sigma}^M_d ) + m {\cal F}^{\rm G}_{\sigma}(d_{\theta}, \sigma_{\theta}; \accentset{\ast}{\sigma}^M_{\sigma} )$, hence saturating \eqref{eq:1mode_bound} with global measurement strategies in the limit of many copies, even if in the single copy case the inequality is strict.

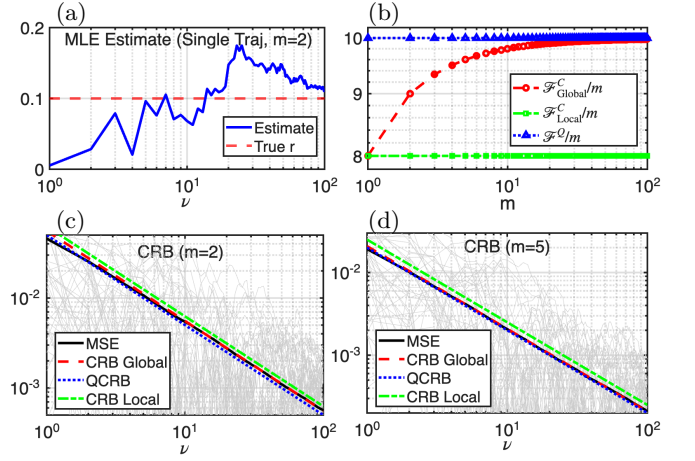
\begin{figure}
    \centering
    % Change 'grid' to 'grid=false' once you are done positioning
    \begin{overpic}[width=1\linewidth, grid=false, tics=10]{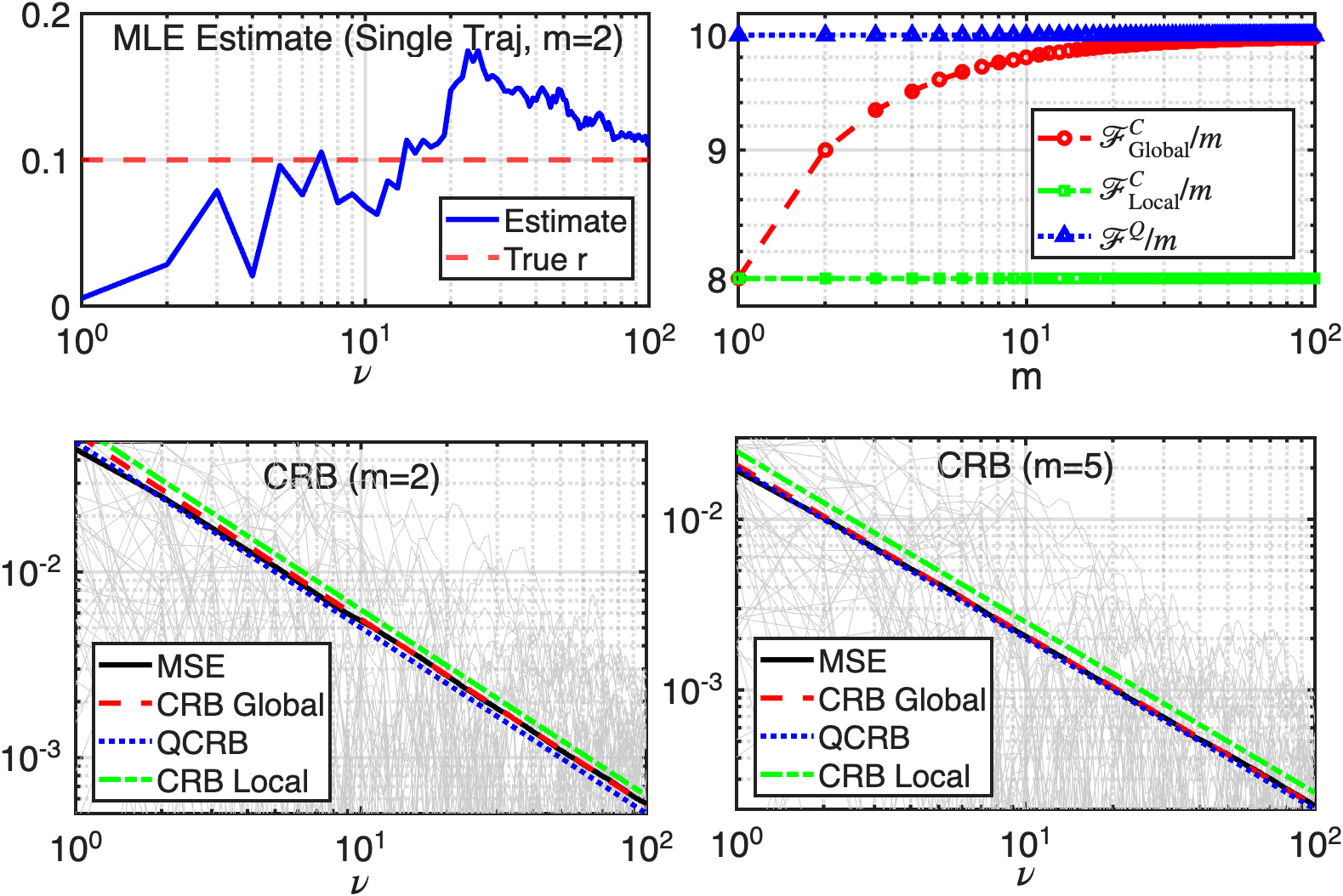}
        \put(7,67){(a)}
        \put(55,67){{(b)}}
        \put(7,35){{(c)}}
        \put(55,35){{(d)}}
    \end{overpic}
    \caption{Monte Carlo simulation of a maximum likelihood estimation (MLE) strategy. (a) Shows how the MLE gets close to the true value with increasing data size $\nu$. (b) Shows how our proposed global strategy can improve local strategies and saturate the QCRB. Panels (c) and (d) show the relevance of the FI as a figure of merit; the MSE by using the MLE very quickly saturates the relevant CRB for two different scenarios with $m=2$ and $m=5$ respectively. The MSE is achieved by averaging the square error over $10^4$ different Monte Carlo simulations---thin gray lines---which shows the CRB is saturated with as few as $\nu=3$ repetitions. Here, we have set $\alpha = \sqrt{2}$, and true $r = 0.1$.}
    \label{fig:MLE}
\end{figure}

{\it Example I: Unitary encoding of squeezing---}We now provide an example where joint Gaussian measurements improve upon local strategies, and even close a significant gap between the optimal Gaussian FI and the QFI. 
The squeezing operation is a single mode symplectic transformation given by 
\begin{align}
    \hspace{-.25cm}d \mapsto \left(\begin{smallmatrix}
        e^{r} & 0 \\
        0 & e^{-r}
    \end{smallmatrix}
    \right) d \eqqcolon d_r,\hspace{.0cm}
    \sigma \mapsto \left(\begin{smallmatrix}
        e^{r} & 0 \\
        0 & e^{-r}
    \end{smallmatrix}
    \right) \sigma \left(\begin{smallmatrix}
        e^{r} & 0 \\
        0 & e^{-r}
    \end{smallmatrix}
    \right) \eqqcolon\sigma_r    .
\end{align}
We are interested in the estimation of $r$ around $r_0=0$.
The input to the channel is a coherent probe state with $d = [\alpha, \alpha]^T$ and $\sigma = I_2$. 
A general pure Gaussian measurement can be seen as a rotated squeezed vacuum with 
\begin{align}
\sigma^{M}(s,\xi) = \left(\begin{smallmatrix}
    \cosh(2s) + \sinh(2s)\cos(\xi) & \sinh(2s) \sin(\xi) \\
    \sinh(2s)\sin(\xi) & \cosh(2s) - \sinh(2s)\cos(\xi)
\end{smallmatrix}\right),
\end{align}
with $s\in\mathbb{R}^+$ and $0\le\xi<2\pi$.
By using Eqs.~\eqref{eq:FI_2terms} one can simply show that
\begin{align}
    {\cal F}_{d}^{\rm G}(d_r,\sigma_r; \sigma^{M})  &=
    2 \alpha^2 (1+\sin{\xi} \tanh{s})\\
    {\cal F}_{\sigma}^{\rm G}( \sigma_r; \sigma^{M})  &=
    1+\cos(2 \xi) \tanh ^2{s}.
\end{align}
Straightforward calculations show that the maximum of the first term is $\accentset{\ast}{\cal F}_{d}^{\rm G}(d_r,\sigma_r) = 4\alpha^2$, achieved by setting $s\rightarrow\infty$ and $\xi = \pi/2$. While, the maximum of the second term is $\accentset{\ast}{\cal F}_{\sigma}^{\rm G}(d_r,\sigma_r) = 2$, achieved by setting $s\rightarrow \infty$ and $\xi \in \{0, \pi\}$. Clearly these two optimal settings are homodyne detections in directions that are not parallel to each other, hence optimizing both terms simultaneously requires incompatible measurements.
Let us now assume $\alpha^2 \geq 2$. Then, one can see that the overall optimal Gaussian measurement---that maximizes the sum of the two terms---is achieved by setting $s\to\infty$ and $\xi = \pi/2$, leading to $\accentset{\ast}{\cal F}^{\rm G}(d_r,\sigma_r) = 4\alpha^2$. 
If we use local Gaussian measurements on $m$ copies of the probe state, the maximum FI is $\accentset{\ast}{\cal F}^{\rm G}(d_r^{\oplus m}, \sigma_r^{\oplus m};{\rm Local})=4\alpha^2m$.
However, the total Fisher information for our proposed global strategy becomes ${\cal F}^{\rm G}(d_r^{\oplus m}, \sigma_r^{\oplus m};{\rm Global}) = 4\alpha^2m+2(m-1)$, which beats the optimal local Gaussian strategy for $m \geq 2$. Let us finally remark that, for our model the QFI can also be calculated, which in the $m$ copy case reads ${\cal F}^Q(d_r^{\oplus m}, \sigma_r^{\oplus m}) = 
4m\alpha^2 + 2m$.
This means that while individual Gaussian measurements result in a maximal FI of $4\alpha^2$ per copy, acting jointly on more copies allows a Gaussian FI of $ 4\alpha^2+2-1/m$ per probe, thus in the limit of many copies reaching the QFI with global Gaussian measurements. In Fig.~\ref{fig:MLE} we depict the MLE of the squeezing parameter, and show that indeed the estimator saturates the CRB with few repetitions, as expected, since the outcome distribution is in the exponential family. See also the Appendix~\ref{App:MLE}.
%%%%%%%%%%%%%%
%%%%%%%%%%%%%%

{\it Example II: Dissipative encoding of losses---}An important channel in quantum optics is the loss channel which describes how the environment around optical fibers can influence the signal they transmit. For certain communication tasks, it is essential to keep losses under control. The channel, can be defined as follows
\begin{align}
    d\mapsto \sqrt{\tau} d \eqqcolon d_{\tau}, \sigma \mapsto \tau\sigma + (1-\tau)\mu I \eqqcolon \sigma_{\tau}. 
\end{align}
Here, $\mu I$ represents the environment's thermal state. At zero temperature $\mu = 1$, but generally at finite temperatures, $\mu > 1$. Our task is the estimation of $\tau$. 

Consider the following input state $d = [0, d_p]$ and $\sigma = {\rm diag}[\sigma_x, \mu]$, which leads to the following output $d_{\tau} = [0, \sqrt{\tau}d_p]$ and $\sigma_{\tau} = {\rm diag}[(1-\tau)\mu+\tau\sigma_x, \mu]$. Let us specifically take a scenario in which $\sigma_x = 2 \mu$, $d_p = \mu = 1$, and $\tau_0 = 1$ (i.e., the channel is close to an ideal channel).

One can show that ${\cal F}^{\rm G}_d(d_{\tau},\sigma_{\tau};\accentset{\ast}{\sigma}_{d}^M) = 1/4$ by using the optimal measurement $\accentset{\ast}{\sigma}_{d}^M = {\rm lim}_{s\to 0}{\rm diag}[1/s, s]$. One can also check that ${\cal F}^{\rm G}_{\sigma}(\sigma_{\tau};\accentset{\ast}{\sigma}_{\sigma}^M) = 1/8$ by using the optimal measurement $\accentset{\ast}{\sigma}_{\sigma}^M = {\rm lim}_{s\to 0}{\rm diag}[s, 1/s]$. Clearly, the two optimal measurements are incompatible. In fact, one can show that the optimal Gaussian measurement is ${\rm lim}_{s\to 0}{\rm diag}[1/s, s]$ leading to $\accentset{\ast}{\cal F}^{\rm G}(d_{\tau},\sigma_{\tau}) = 1/4 < 1/4+1/8 = {\cal F}^{\rm G}_d(d_{\tau},\sigma_{\tau};\accentset{\ast}{\sigma}_{d}^M) + {\cal F}^{\rm G}_{\sigma}(\sigma_{\tau};\accentset{\ast}{\sigma}_{\sigma}^M)$. Again, our proposed joint strategy allows for saturating this bound at the limit of many copies, i.e., ${\cal F}^{\rm G}(d_r^{\oplus m}, \sigma_r^{\oplus m};{\rm Global}) = m/4 + (m-1)/8$. Contrary to the previous example, in this case we have ${\cal F}^Q(d_{\tau},\sigma_{\tau}) = 1/4+1/6$ and the gap with Gaussian measurements cannot be closed using our proposed global Gaussian measurement.

{\it Discussion---}In this work, we focused on the Fisher information of\textit{ Gaussian measurements}. To this end, we fixed the input state as a Gaussian state to isolate the specific role of the measurement. Similar questions may be asked when an extra optimization over Gaussian input states is considered, although in this case one should further take energy or maximum photon number constraints into account~\cite{PhysRevResearch.4.033092,PRXQuantum.6.020351}. We proved that if information about a parameter is encoded in either the displacement or the covariance matrix of a Gaussian system, the FI of the optimal Gaussian measurement remains additive. This implies that there is no advantage in joint Gaussian measurements. We also proposed a specific strategy based on feasible elements of linear optics (beam splitters) that leads to super additivity of the optimal Gaussian FI when information is encoded in both moments. Our proposed Gaussian measurement is indeed asymptotically optimal; this is not necessarily the case for finite $m$. It is still an open problem to find \textit{the} optimal global Gaussian strategies in the limit of few copies.
This becomes relevant in a Bayesian scenario, where optimal strategies can differ from the optimal frequentist strategies~\cite{morelli2021bayesian,Ravell_Morelli_2025}.
Lastly, it is known that in the presence of uncorrelated noise the scaling described by the QCRB persists only up to a certain point~\cite{DemkowiczDobrzanskiKolodynskiGuta2012, EscherDavidovichZaguryDeMatosFilho2012, ChavesBraskMarkiewiczKolodynskiAcin2013} and one needs additional resources to combat noise~\cite{SekatskiSkotiniotisDuer2016, SekatskiSkotiniotisKolodynskiDuer2017}. Here, we focused on Gaussian measurements on uncorrelated copies of a states, it remains an interesting direction to study protocols where only noisy Gaussian measurements are available.

{\bf Acknowledgments---}We thank Alexssandre de Oliveira Junior for discussions on early stages of the work.
The authors acknowledge TU Wien Bibliothek for financial support through its Open Access Funding Programme.
This research was funded in part by the Austrian Science Fund (FWF) [“NOQUS,” Grant DOI: 10.55776/PAT1969224],
the Austrian Federal Ministry of Education, Science, and Research via the Austrian Research Promotion Agency (FFG) through the project HDcode funded by the European Union - NextGenerationEU. Support from OpenSuperQ+100 (Grant No. 101113946) of the EU Flagship on Quantum Technologies, Project Grant No. PID2024-156808NB-I00; Spanish Ramón y Cajal Grant No. RYC-2020-030503-I by MICIU/AEI/10.13039/501100011033, “ERDF A way of making Europe” and “ERDF Invest in your Future”,
from the Spanish Ministry for Digital Transformation and of Civil Service of the Spanish Government through the QUANTUM ENIA project call-Quantum Spain, and by the EU through the Recovery, Transformation and Resilience Plan--NextGenerationEU within the framework of the Digital Spain 2026 Agenda, 
and from Basque Government through Grant No. IT1470-22,
and the Elkartek project KUBIBIT - kuantikaren berrikuntzarako
ibilbide teknologikoak (ELKARTEK25/79). We acknowledge funding from Basque Government through Grant No. IT1470-22 and the IKUR Strategy under the collaboration agreement between Ikerbasque Foundation and BCAM on behalf of the Department of Education of the Basque Government. 

%%%%%%%%%%%
\bibliographystyle{bibstyle}
\bibliography{references}
\newpage
\appendix
\onecolumngrid
\section{Further details of the \textit{Result 1}}\label{app:proof}
\subsection{Information only on the displacement vector}
Take an $N$ mode system with the covariance matrix $\sigma = \oplus_{i=1}^N\mu_{i} S S^T$. Furthermore, let's denote $d^{\prime} = S^{-1} \partial_{\theta} d_{\theta} = \oplus_{i=1}^N d_i^{\prime}$. We have ${\cal F}^{\rm Q}(d_{\theta},\sigma) = \sum_{i=1}^N|d^{\prime}_i|^2/\mu_i$. On the other hand, the FI for an arbitrary Gaussian measurement $\sigma^M = S \gamma^{M} S^T$ reads
\begin{align}
    {\cal F}^{\rm G}(d_{\theta}, \sigma; \sigma^M) & = d^{\prime T} (\oplus_{i=1}^N \mu_i I + \gamma^M)^{-1} d^{\prime}.
\end{align}
Take the following ansatz $\gamma^M = \oplus_{i=1}^N O_i^T Z O_i$ with $O_i$ being a rotation, and $Z = \lim_{z\to 0} {\rm diag} (z,1/z)$. For this choice, we have
\begin{align}
    {\cal F}^{\rm G}(d_{\theta}, \sigma; \sigma^M) & = \sum_{i=1}^N d_i^{\prime T} O^T_i (\mu_i I + Z)^{-1} O_i d_i^{\prime} = \sum_{i=1}^N\frac{([O_id_i^{\prime}]_{11})^2}{\mu_i}.
\end{align}
One can always choose $O_i$ such that $O_i d_i^{\prime} = {\rm diag}(|d_i^{\prime}|, 0)$. Thus we found a Gaussian measurement whose FI is equal to the QFI. 
\subsection{Information only on the covariance matrix}
The proof is mainly discussed in the main text. Here we prove Eq.~\eqref{eq:ineq2}.
We take $\{{W_i}^2\}$ to be the set of eigenvalues of ${W}^2$ sorted descendingly, i.e., $ W_i^2\geq W_j^2$ if $i<j$. Similarly take the eigenvalues of $V^2$, to be $\{V_i^2\}$ descendingly ordered and $O_G$ to be an orthogonal matrix. In the main text, we claimed
\begin{align}
      \operatorname{Tr} \left[ V^2  \left(O_G^T W^2O_G \right)  \right] \leq \sum_i V_i^2W_i^2.
\end{align}
\begin{proof}
Let $M = O_G^T W^2 O_G=O_G^{\prime T} W^2_D O_G^\prime$ where $W^2_D={\rm diag}\{W_1^2,W_2^2,\dots,W_{2m}^2\}$. Since the matrix $p_{ij}= |O_{ij}|^2$ is a doubly stochastic matrix---$ p_{ij}\geq0$, $\sum_i p_{ij}=\sum_j p_{ij}=1$---the sequence $M_i= \sum_j |O_{ij}|^2W_j^2$ is majorized by the sequence $W_i$, that is to say $\{M_i\}\preceq \{W_i^2\}$ \cite{bhatia2007perturbation}. We introduce the function $ f(\{M_i\})$,
\begin{align}
      f(\{M_i\})\coloneqq \sum_i V_i^2M_i \equiv \operatorname{Tr} \left[ V^2  M  \right].
\end{align}
Such real valued function $f(\{M_i\})$ is \textit{Schur-convex} if $\{M_i\}\preceq \{M^{\prime}_i\}$ implies that $f(\{M_i\}) \leq f(\{M^{\prime}_i\})$ \cite{marshall1979inequalities}. One can further show that $f$ is Schur-convex if and only if for a given $\{M_i\}$ in decreasing order, $M_i\geq M_{i+1}$, it is satisfied that $f(\{M_1,\dots,M_k+\epsilon,M_{k+1}-\epsilon,\dots, M_{2m}\})$ is non-decreasing in $\epsilon$ over the region
\begin{align*}
    &0\leq \epsilon\leq \min[M_{k-1}-M_k,M_{k+1}-M_{k+2}]\hspace{1.5cm} k=1,\dots,2m-2,\\
    &0\leq \epsilon\leq M_{2k-2}-M_{2k-1}\hspace{4cm} k=2m-1,
\end{align*}
for any chosen $k$. 

In our specific problem, we have
\begin{align*}
    \partial_\epsilon f(\{M_1,\dots,M_k+\epsilon,M_{k+1}-\epsilon,\dots, M_{2m}\})=  \partial_\epsilon[V_k^2(M_k+\epsilon)+V_{k+1}^2(M_{k+1}-\epsilon)]=V_{k}^2-V_{k+1}^2\geq0,
\end{align*}
where in the last equality we have used that $\{V_i^2\}$ is descendingly ordered. Since $\{M_i\}$ and $\{W_i^2\}$ are two  descendingly ordered sets satisfying $\{M_i\}\preceq \{W_i^2\}$ and $f(\{M_i\} $ is Schur-convex it follows  $ f(\{M_i\})\leq f(\{W_i^2\})$ which proves the claim.
\end{proof}

\section{Extension to isothermal states with higher number of modes}\label{app:IsothermalCov}
Here, we prove that $ \accentset{\ast}{\cal F}^{\rm G}(d_{\theta}^{\oplus m}\sigma^{\oplus m}_\theta) =m\accentset{\ast}{\cal F}^{\rm G}(d_{\theta},\sigma_\theta) $ for isothermal models with higher ($N$) number of modes and vanishing displacement contribution ($\partial_{\theta} d_{\theta}=0$). For any measurement $\sigma^{M}$ we have
\begin{align*}
      {\cal F}^{\rm G}(d^{\oplus m},\sigma^{\oplus m}_\theta;\sigma^{M})   =  \frac{1}{2}\operatorname{Tr} \left[\left((\sigma^{\oplus m}_\theta +  \sigma^{M})^{-1} \partial_{\theta}\sigma^{\oplus n}_\theta\right)^2 \right]=\frac{1}{2}\operatorname{Tr} \left[\left((\mu I +  \gamma^{M})^{-1} [S^{-1}\partial_{\theta}{\sigma}S^{-T}]^{\oplus m}\right)^2 \right],
\end{align*}
where we used the Williamson decomposition of the isothermal model $\sigma_\theta = \mu SS^T $ and defined $\gamma^{M}=S^{-1}\sigma^{M}S^{-T} $.
Any covariance matrix can be diagonalized by an orthosymplectic $O_G$ , 
\begin{align*}
   {\gamma}_s^{M} &= O_G\left(\oplus_{i=1}Z_i\right) O_G^T,\hspace{.5cm} z\in(0,1],
\end{align*}
with $Z_i = {\rm diag}(z_i, 1/z_i)$.
Here, the sub-index $G$ emphasizes that in general it is a global orthogonal transformation. We note that $(\mu I + \oplus_iZ_i)^{-1}=\oplus_i^m{\rm diag}\{\frac{1}{\mu+z_1},\frac{z_1}{\mu z_1+1},\frac{1}{\mu+z_2},\dots,\frac{z_N}{\mu z_N+1} \}\eqqcolon V$. By further defining $W\coloneqq [S^{-1}\partial_{\theta}{\sigma}S^{-T}]^{\oplus m}$, we have
\begin{align}
    {\cal F}^{\rm G}(d^{\oplus m}, \sigma^{\oplus m}; \sigma^{M})  =\frac{1}{2}\operatorname{Tr} \left[\left(V\: O_G^T WO_G\right)^2 \right]\nonumber \leq \frac{1}{2}\operatorname{Tr} \left[ V^2 O_G^T W^2 O_G  \right],
\end{align}
where we have used the property derived in the main text, ${\rm Tr} [(VO_G^T W O_G )^2] \leq {\rm Tr} [V^2O_G^T W^2 O_G ]$ . This inequality is tight if $[V, O_G W O_G] = 0$, that is, $V$ and $O_G W O_G^T$ can be diagonalized in the same basis.
We will show in the Appendix below that for isothermal models it always exists an orthosymplectic transformation $O'$ that diagonalizes $W$, 
\begin{align}\label{eq:W_D_orthosymplectic}
O^{\prime} W^2 O^{\prime T}=  W_D^2=& \left(\begin{smallmatrix}
   W_1^2&0&0&0&0\\
   0&W_2^2&0&0&0\\
    0&0&W_3^2&0&0\\
    0&0&0&\ddots&0\\
    0&0&0&0&W_{2N}^2
   \end{smallmatrix} \right)^{\oplus m }, \hspace{1cm} W_{2j}^{2}\leq W_{2j-1}^{2} {~~\rm with~~}1\leq j \leq N.
\end{align}
By defining $ {\tilde O}_G=O^\prime O_G$ we have
\begin{align}
 {\cal F}^{\rm G}(d^{\oplus m}, \sigma^{\oplus m}; \sigma^{M}) \leq\frac{1}{2}\operatorname{Tr} \left[ V^2 O_G^T W^2 O_G  \right]=  \frac{1}{2}\operatorname{Tr} \left[ V^2 {\tilde O^T}_G W^2_D {\tilde O}_G  \right]= \frac{1}{2}\sum_{j=1}^{2 m N}V^2_{jj} ( {\tilde O^T}W_D^2  {\tilde O})_{jj}=\frac{1}{2}\sum_{i,j =1}^{2mN} {\tilde O}_{ij}^2 V^2_{jj} W^2_{i}.
 \label{eq:B2}
\end{align}
Note that due to the block-diagonal nature of $W_D$ its eigenvalues are repeated $m$ times with peridicity $2N$, i.e. $W_i=W_{i+2N}$.
The symplectic nature of an orthosymplectic matrix ${\tilde O}$ impose additional relations among its elements, namely $({\tilde O}_{2i-1,2j})^2=({\tilde O}_{2i,2j-1})^2$ and $({\tilde O}_{2i-1,2j-1})^2=({\tilde O}_{2i,2j})^2$. We expand Eq.~\eqref{eq:B2} in odd and even indices,
\begin{align*}
    {\cal F}^{\rm G}(d^{\oplus m}, \sigma^{\oplus m}; \sigma^{M}) & \leq\frac{1}{2}\sum_{ij}{\tilde O}_{ij}^2V^2_{jj} W^2_{i}  =   \frac{1}{2}\sum_{i=1}^{2mN}\sum_{j=1}^{mN}V^2_{2j-1,2j-1} {\tilde O}_{i,2j-1}^2W^2_{i}+V^2_{2j,2j}  {\tilde O}_{i,2j}^2W^2_{i}\\
    & = \frac{1}{2} \sum_{i,j=1}^{mN}V^2_{2j-1,2j-1} {\tilde O}_{2i-1,2j-1}^2W^2_{2i-1}+V^2_{2j-1,2j-1} {\tilde O}_{2i,2j-1}^2W^2_{2i}\\
    &\hspace{1.1cm}+V^2_{2j,2j} {\tilde O}_{2i-1,2j}^2W^2_{2i-1}+V^2_{2j,2j} {\tilde O}_{2i,2j}^2W^2_{2i}\\
    & =\frac{1}{2} \sum_{i,j=1}^{mN} {\tilde O}_{2i,2j}^2 (V^2_{2j-1,2j-1} W^2_{2i-1}+V^2_{2j,2j} W^2_{2i})+ {\tilde O}_{2i,2j-1}^2 (V^2_{2j,2j} W^2_{2i-1}+V^2_{2j-1,2j-1} W^2_{2i}).
\end{align*}
Noting that $ W_{2i}^2\leq W_{2i-1}^2$---see Eq.~\eqref{eq:W_D_orthosymplectic}---and that $ V_{2j,2j}^2 = \frac{z_j^2}{(z_j\mu+1)^2}\leq\frac{1}{(\mu+z_j) ^2}=V_{2j-1,2j-1}^2$ $\forall z_j \in (0, 1]$,  and using the rearrangement inequality, we get
\begin{align*}
  {\cal F}^{\rm G}(d^{\oplus m}, \sigma^{\oplus m}; \sigma^{M}) 
    & \leq \frac{1}{2} \sum_{i,j=1}^{mN} ({\tilde O}_{2i,2j}^2+{\tilde O}_{2i,2j-1}^2 ) (V^2_{2j-1,2j-1} W^2_{2i-1}+V^2_{2j,2j} W^2_{2i})\\
     &\leq \frac{1}{2}  \sum_{i,j=1}^{mN} ({\tilde O}_{2i,2j}^2+{\tilde O}_{2i,2j-1}^2 ) \max_{z_j}\{V^2_{2j-1,2j-1} W^2_{2i-1}+V^2_{2j,2j} W^2_{2i}\}.
\end{align*}
To proceed further note that $\max_{z_j}\{V^2_{2j-1,2j-1} W^2_{2i-1}+V^2_{2j,2j} W^2_{2i}\}$ is independent of $j$ and only depends on the $i$ index via $W^2_{2i-1} $ and $ W^2_{2i}$ so we can write
\begin{align*}
     {\cal F}^{\rm G}(d^{\oplus m}, \sigma^{\oplus m}; \sigma^{M}) & 
      \leq \frac{1}{2}\sum_{j=1}^{mN}({\tilde O}_{2i,2j}^2+{\tilde O}_{2i,2j-1}^2 )\sum_{i=1}^{mN} \max_{z_j}\{V^2_{2j-1,2j-1} W^2_{2i-1}+V^2_{2j,2j} W^2_{2i}\} \\
      & =\frac{1}{2} \sum_{i=1}^{mN} \max_{z_j}\{V^2_{2j-1,2j-1} W^2_{2i-1}+V^2_{2j,2j} W^2_{2i}\}\\
      & =\frac{m}{2} \sum_{i=1}^{N} \max_{z_j}\{V^2_{2j-1,2j-1} W^2_{2i-1}+V^2_{2j,2j} W^2_{2i}\},
\end{align*}
where in the last two lines we have used $\sum_{j=1}^{mN}({\tilde O}_{2i,2j}^2+{\tilde O}_{2i,2j-1}^2 )=1$ and the periodicity of the diagonal elements $\{W_i\}$, respectively. The final expression is $m$ times the optimal Fisher information of a single copy. Therefore, we proved that the Fisher information of any global measurement is bounded by the  Fisher information of the optimal local measurement.

\subsection{More details on isothermal models and the proof of Eq.~\eqref{eq:W_D_orthosymplectic}}\label{app:Isothermal}
In the following we prove that for isothermal models it does always exists an orthosymplectic matrix that diagonalizes $W = S^{-1}\partial_{\theta}{\sigma}_{\theta}S^{-T}$, that is Eq.~\eqref{eq:W_D_orthosymplectic}.
Recall that if the covariance matrix of a system can be written as $\sigma_\theta=\mu SS^T $ it is called an \textit{isothermal model}. We note that isothermal models include all pure models \cite{monras2013phase} as well as some relevant mixed models \cite{blandino2012homodyne}.  To begin with, note that one can always write $W = A + \alpha I$ with $A$ a Hamiltonian matrix---i.e., satisfying $A^T \Omega + \Omega A = 0$---and $\alpha$ a scalar:
    \begin{align*}
        W& = S^{-1}\partial_\theta \sigma_\theta S^{-T}
        =\mu S^{-1} (\partial_{\theta}S) + \mu  (\partial_{\theta}S^T) S^{-T}+ (\partial_{\theta}\mu) I  
    =\mu(M^T+ M) +(\mu^{\prime}) I,
    \end{align*}
    with $\mu^{\prime}\partial_{\theta}\mu$ and  $ M \coloneqq S^{-1}\partial_{\theta} S$ being a Hamiltonian matrix, as can be simply checked by using $\partial_{\theta} (S\Omega S^T = \Omega)$. Note that if $M$ is Hamiltonian matrix, so is $M^T$ as well as $\mu(M^T+ M)$.

Furthermore, since $\mu (M+M^T)$ is symmetric and Hamiltonian there exists an orthosymplectic transformation $O$ that diagonalizes it---see below for a proof. Therefore, $O$ also diagonalizes $W$, and one can write
\begin{align}\label{eq:W_diag_orthosym}
    O W^2 O^T = {\rm diag}\{(\mu'+\sigma_1)^2,(\mu'-\sigma_1)^2, \dots , (\mu'+\sigma_N)^2, (\mu'-\sigma_N)^2 \}.
\end{align}
with $\sigma_j \geq 0~\forall j$.
Permutations between the quadratures of a given mode keep the matrix diagonal. So we define the single-mode orthosymplectic permutation matrix $O_p$ as,
\begin{align}
     O_p {\rm diag}((\mu'+\sigma_j)^2,(\mu'-\sigma_j)^2) O_p^T & = {\rm diag}((\mu'+\sigma_j)^2,(\mu'-\sigma_j)^2) ~~\textit{if}\quad \mu'\geq0 , \nonumber\\ 
    O_p {\rm diag}((\mu'+\sigma_j)^2,(\mu'-\sigma_j)^2) O_p^T & = {\rm diag}((\mu'-\sigma_j)^2,(\mu'+\sigma_j)^2) ~~\textit{if}\quad \mu'\leq0 .\label{eq:B4}
\end{align}
To conclude the matrix $W^2$ can be always diagonalized by the orthosymplectic transformation  $O^{\prime} = O_p^{\oplus N}O$,
\begin{align}
    O^{\prime} W^2 O^{\prime T}={\rm diag}\{W_1^2,W_2^2, \dots ,W_{2N-1}, W_{2N}^2 \}\label{eq:B5},
\end{align}
where $ W_{2i}^2\leq W_{2i-1}^2$ because of Eq.~\eqref{eq:B4}.

\subsection{Proof of Eq.~\eqref{eq:W_diag_orthosym}}We now show that any real matrix $A$ which is both symmetric and Hamiltonian can be diagonalised by a matrix that is both orthogonal and symplectic (orthosymplectic). It is more convenient to change to the convention given by $R = [q_1, \dots,q_N,p_1,\dots, p_N]^T$.
Write $A$ in \(N\times N\) blocks,
\begin{align*}
  A=\begin{pmatrix} P & Q \\ R & S \end{pmatrix}.  
\end{align*}
Symmetry implies
\begin{align*}
P^T=P,\qquad S^T=S,\qquad R^T=Q.
\end{align*}
The Hamiltonian condition \(A^T \Omega + \Omega A = 0\) gives, after a short block computation,
\begin{align*}
S^T=-P,\qquad Q^T=Q,\qquad R^T=R.
\end{align*}
Combining both sets of relations, we obtain the general form
\begin{align*}
A=\begin{pmatrix} P & Q \\ Q & -P \end{pmatrix},
\qquad P^T=P,\; Q^T=Q.
\end{align*}
The symplectic orthogonal group of $2N\times2N$ real matrices is isomorphic to complex unitaries $U(N)$ \cite{serafini2023quantum}. We identify \(\mathbb{R}^{2N}\simeq\mathbb{C}^N\) via
\((x,y)\mapsto x+i y\).
Then
\begin{align*}
A=\begin{pmatrix} P & Q \\ Q & -P \end{pmatrix}
\quad\longleftrightarrow\quad
H = P + iQ.
\end{align*}
where $H$ is $N\times N$ complex symmetric matrix because both \(P\) and \(Q\) are symmetric,
\begin{align*}
H^T = P^T + i Q^T = P + i Q = H.
\end{align*}
Takagi decomposition \cite{horn2012matrix,houde2024matrix} states that every complex symmetric matrix admits a unitary decomposition so that \begin{align}
    H = U\,\Sigma\,U^T
\end{align}
where $U$ is unitary and
\(\Sigma=\operatorname{diag}(\sigma_1,\dots,\sigma_N)\) with
\(\sigma_j\ge 0\).
Write \(U = U_R + i U_I\) with real and imaginary parts \(U_R,U_I\in\mathbb{R}^{N\times N}\), and define
\begin{align}
    O^T =
\begin{pmatrix}
U_R & U_I\\[4pt]
-U_I & U_R
\end{pmatrix}.
\end{align}

Then one can directly check that \(O\in O(2N)\cap Sp(2N,\mathbb{R})\), i.e.\ it is both orthogonal and symplectic. Furthermore, a direct block computation shows that
\[
O A O^T =
\begin{pmatrix}
\Re(U^T H U) & \Im(U^T H U)\\[4pt]
\Im(U^T H U) & -\Re(U^T H U)
\end{pmatrix}.
\]
Substituting $U^T H U = \Sigma$ into the expression above yields
\begin{align}
    O A O^T =
\begin{pmatrix}
\Sigma & 0\\[4pt]
0 & -\Sigma
\end{pmatrix},
\end{align}
which is diagonal.
If we now go back to the original convention of $R=[x_1,p_1,\dots,x_N,p_N]$, and use $W= A + \mu^{\prime}I$, we have
\begin{align*}
    O W O^T & = {\rm diag}\{(\mu'+\sigma_1),(\mu'-\sigma_1), \dots , (\mu'+\sigma_N), (\mu'-\sigma_N) \}.
\end{align*}

\section{The maximum likelihood estimator for the squeezing estimation example}\label{App:MLE}
In the squeezing estimation problem, the Gaussian states are measured in blocks of $m$ (jointly). To generate more data, one repeats the process $\nu$ times. An estimator is needed to process the data collected from all $m\nu$ copies, which we chose to be the maximum likelihood estimator. 

In our joint strategy, after shifting all the displacement to one of the copies, say the $m$th one, that one is measured in a specific quadrature, while the first $m-1$ are measured in another direction.
We repeat this a total of $\nu$ times.

If we denote the outcomes of all $\nu$ measurements on the $j$th mode with ${\vec x}^{(j)} = ({x}^{(j)}_1,\dots, {x}^{(j)}_{\nu})$, then by using Eq.~\eqref{eq:Gaussian_meas_pdf} for the measurement that we introduced in the main text, the likelihood reads
\begin{gather}
    p({\vec x}^{(1)}, \dots {\vec x}^{(m)} | r) 
     = \Pi_{k=1}^{\nu} p({ x}^{(1)}_k, \dots {x}^{(m)}_k | r) 
    %%%%%
    = \Pi_{k=1}^{\nu} \frac{e^{\frac{-\sum_{j=1}^{m-1}\left[x_k^{(j)}\right]^2}{2e^{2r}}}}{\left(2\pi e^{2r}\right)^{\frac{m-1}{2}}} \times \frac{e^{\frac{-\left[\sqrt{m}\alpha(e^r - e^{-r}) - x_k^{(m)}\right]^2}{e^{2r} + e^{-2r}}}}{\sqrt{\pi (e^{2r} + e^{-2r})}}\\
    %%%%%%
    %%%%%%
     = 
    \left(2\pi e^{2r}\right)^{-\frac{\nu(m-1)}{2}} {\rm Exp}\left[\frac{-\sum_{k=1}^{\nu}\sum_{j=1}^{m-1}\left[x_k^{(j)}\right]^2}{2e^{2r}}\right] \times \left({\pi (e^{2r} + e^{-2r})}\right)^{-\frac{\nu}{2}} {\rm Exp}\left[\frac{-\sum_{k=1}^{\nu}\left[\sqrt{m}\alpha(e^r - e^{-r}) - x_k^{(m)}\right]^2}{e^{2r} + e^{-2r}}\right].\nonumber
\end{gather}
The log-likelihood reads
\begin{align}
    \log p({\vec x}^{(1)}, \dots {\vec x}^{(m)} | r) & = -\sum_{k=1}^{\nu} \left[ \frac{\sum_{j=1}^{m-1} \left[x_k^{(j)}\right]^2}{2e^{2r}} + \frac{\left[\sqrt{m}\alpha(e^r - e^{-r}) - x_k^{(m)}\right]^2}{e^{2r} + e^{-2r}} \right] - \nu(m-1) r - \frac{\nu}{2} \log(2\cosh(2r)) + {\rm cst.}
\end{align}
The maximum likelihood estimator can be numerically found as the value of $r$ that maximises the above quantity for the observed data. A MATLAB code is supplemented that simulates the outcomes, and then performs this maximization~\cite{Mehboudi_Fully_Gaussian_metrology_2025}.

\end{document}